\documentclass{aa}
\usepackage{graphicx}
\usepackage{natbib}
\def\be{\begin{equation}}
\def\ee{\end{equation}}

\def\EB{\hbox{${\rm {\bf E} \times {\bf B}}$}}

\begin{document}

\title{Interrelation between radio and X-ray signatures of drifting subpulses in pulsars}
%\subtitle{X-ray emission and drifting subpulses}
\author{ Janusz Gil,\inst{1,2}
%\thanks{E-mail: jag@astro.ia.uz.zgora.pl}
\and
George Melikidze,\inst{1,3}
\and
Bing Zhang\inst{2}}
\institute{Institute of Astronomy, University of Zielona G\'ora,
Lubuska 2, 65-265, Zielona G\'ora, Poland \\ \and Department of Physics, University of Nevada, Las Vegas, USA\\ \and Abastumani
Astrophysical Observatory, Al. Kazbegi ave. 2a, 0160, Tbilisi, Georgia }
\date{ }

\abstract
% context heading (optional) {} Leave it empty if necessary
{}
% aims heading (mandatory)
{We examined a model of partially screened gap region above the polar cap, in which the electron-positron plasma
generated by sparking discharges coexists with thermionic flow ejected by the bombardment of the surface beneath these
sparks. Our special interest was the polar cap heating rate and the subpulse drifting rate, both phenomena presumably
associated with sparks operating at the polar cap.}
% methods heading (mandatory)
{We investigated correlation between the intrinsic drift rate and polar cap heating rate and found that they are
coupled to each other in such a way that the thermal  X-ray luminosity $L_x$ from heated polar cap depends only on the
observational tertiary subpulse drift periodicity $\hat{P}_3$ (polar cap carousel time).}
% results heading (mandatory)
{Within our model of partially screened gap we  derived the simple formula relating $L_x$ and $\hat{P}_3$, and showed
that it holds for PSRs B0943$+$10 and B1133+16, which are the only two pulsars in which both $L_x$ and $\hat{P}_3$ are
presently known.}
% conclusions (optional)
{}

\maketitle
\section{Introduction}

Revealing the problem of the drifting subpulses phenomenon seems to be a keystone for the understanding of the pulsar
radio emission mechanism. In the model of \citet[][ RS75 henceforth]{rs75} the spark-associated subbeams of subpulse
emission circulate around the magnetic axis due to $\mathbf{E}\times\mathbf{B}$ drift of spark plasma filaments. This
model is widely regarded as the most natural and plausible explanation of drifting subpulse phenomenon, at least
qualitatively. Despite its popularity the RS75 model is known to suffer from the so-called binding energy problem
\citep[for review see][]{um96}. In fact, the cohesive energy of surface iron ions were largely overestimated and the
"vacuum gap" envisioned by RS75 was impossible to form. \citet{gm01} and \citet{gm02} revisited the binding energy
problem and concluded that the formation of RS75 gap was, in principle, possible, although it required a very strong
surface magnetic field, much stronger than the dipolar component inferred from the observed spin-down rate. Recent
calculations of \citet{ml06} showed that iron chains are strongly bound in magnetic field close to $10^{14}$ G. This
can only occur in actual pulsars if their surface field is dominated by strong non-dipolar components. Growing evidence
of such non-dipolar surface anomalies accumulates in the literature, both observational and theoretical \citep[for
short review see e.g. ][]{ug04}.  Therefore, it seems natural to assume that the magnetic field at the stellar surface
is dominated by the crust anchored small scale (sun-spot like) components, which can significantly alter the physical
conditions for plasma production in the polar cap region \citep[see][for details]{gmm02}.

Even if the RS75 gap was possible to form, it would not automatically solve the mystery of drifting subpulses, since it
predicts too fast a drifting rate \cite[e.g.][]{dr99}. Motivated by this issue \citet[][ GMG03 henceforth]{gmg03}
developed further the idea of the inner acceleration region (IAR henceforth) above the polar cap by including the
partial screening due to thermionic ions flow from the surface heated by sparks. We will call this kind of IAR the
"partially screened gap" (PSG henceforth). The slow drift corresponds to \EB\ drift of spark generated
electron-positron pairs, until the total charge density within the PSG reaches the co-rotational value. Since the PSG
potential drop is much lower than in the RS75 model, the intrinsic drift rate is compatible with observations (for
details see GMG03).

The PSG model allows a relatively high heating rate of the polar cap surface, compatible with observations, in contrast
to RS75 gap, which produces too much heat. On the other hand, the alternative Space Charge Limited Flow model
\citep{as79} predicts a much lower heating rate \citep[e.g.][]{zh00}. Thus, measuring the thermal X-ray luminosity from
heated polar caps can potentially reveal the nature of IAR. This can help us to understand a mechanism of drifting
subpulses, which appears to be a common phenomenon in radio pulsars. Recently \citet[][WES06 henceforth]{wes06}
presented results of a systematic, unbiased search and found that the fraction of pulsars showing drifting subpulses is
at least 55~\%. They concluded that the conditions for drifting mechanism to work cannot be very different from the
emission mechanism of radio pulsars.

In order to test different available models of IAR in pulsars \citet[][ ZSP05 henceforth]{zsp05} observed the best
studied drifting subpulse radio pulsar PSR B0943$+$10 with the {\em XMM-Newton}. Their observations were consistent
with PSG formed in strong, non-dipolar magnetic field just above the surface of very small polar cap. Recently,
\citet[][KPG06 henceforth]{kpg06} observed the X-ray emission from the nearby PSR B1133$+$16 and found that this case
is also consistent with the thermal radiation from a small hot spot (again much smaller than the canonical polar cap).
PSR B1133$+$16 is almost a twin of PSR B0943$+$10 in terms of $P$ and $\dot{P}$ values and, interestingly, both these
pulsars have very similar X-ray signatures, in agreement with our PSG model (see Table 1).

\begin{table*}
\begin{minipage}{170mm}
\begin{center}

\fontsize{10}{10pt}\selectfont

\caption{Comparison of observed and predicted parameters of thermal emission from hot polar caps}
\begin{tabular}{c c c c c c c c c c c}
\hline \hline Name & $P_3/P$ & \multicolumn{2}{c}{$\hat{P}_{3}/P$} &
\multicolumn{2}{c}{${L_{\mathrm{x}}}/{\dot{E}}\times 10^{3}$} & $b$ & $T_{s}^{\mathrm{(obs)}}$ &
$T_{s}^{\mathrm{(pred)} }$ & $B_{\mathrm{d}}$ & $B_{\mathrm{s}}$ \\
\hline \ PSR B & Obs. & Obs. & Pred. & Obs. & Pred. &
${A_{\mathrm{pc}}}/{A_{\mathrm{bol}}}$ & $10^{6}\ $K & $10^{6}\ $K & $10^{12}$G & $10^{14}$G \\
\hline $0943+10$& 1.86 & $37.4$ & $36_{-2}^{+8}$ & $0.49_{-0.16}^{+0.06}$ & $0.45$ & $60_{-48}^{+140}$ &
$3.1_{-1.1}^{+0.9}$ & $3.3_{-1.1}^{+1.2}$ & $3.95$ & $ 2.37_{-1.90}^{+5.53}$
\\
$1133+16$ & $3^{+2}_{-2}$ & ($33_{-3}^{+3}$) & $27_{-2}^{+5}$ & $0.77_{-0.15}^{+0.13}$ & $0.58_{-0.09}^{+0.12}$  &
$11.1_{-5.6}^{+16.6}$ & $2.8_{-1.2}^{+1.2}$ & $2.1_{-0.4}^{+0.5}$ &
$4.25$ & $ 0.47_{-0.24}^{+0.71}$ \\
\hline
\end{tabular}
\end{center}
\end{minipage}
\end{table*}

In this paper we generalized the treatment of ZSP05 and developed a model for thermal X-ray emission from radio
drifting pulsars. The model matches the observations of PSRs B0943+10 and B1133+16, both in radio and X-rays. There is
a number of pulsars with measured thermal X-ray radiation from small hot polar cap but unfortunately their drifting
properties are not yet known. This is likely to change in the near future due to new sophisticated methods for analysis
of intensity fluctuations in weak pulsars being developed (e.g. WRS06 and references therein).

\section{Charge depleted inner acceleration region}

The IAR above the polar cap results from the deviation of a local charge density $\rho$ from the co-rotational charge
density \citep[][ GJ69 henceforth]{gj69} $\rho_{\rm GJ}=-{\mathbf\Omega}\cdot{\bf B}_s/{2\pi c}\approx{B_s}/{cP}$. As
already mentioned, growing evidence appears, both observationally and theoretically, that the actual surface magnetic
field $B_s$ is highly non-dipolar. One can introduce a scaling factor $b=B_s/B_d$ to describe its magnitude
\citep[][GS00 henceforth]{gs00}, where the enhancement coefficient $b>1$ and $B_d=2\times 10^{12}(P\dot{P}_{-15})^{1/2}
{\rm G}$ is the canonical, star centered dipolar magnetic field, $P$ is the pulsar period and
$\dot{P}_{-15}=\dot{P}/10^{-15}$ is the period derivative.

The polar cap is defined as the locus of magnetic field lines that penetrate the so-called light cylinder.
Conventionally, the polar cap radius $r_{pc}=1.45\times 10^4P^{-0.5}~{\rm cm}$, and its surface area $A_{pc} \sim
2\times 10^8 P^{-1}$ cm$^2$. In the case of non-dipolar surface field the polar cap area must shrink due to flux
conservation of the open field lines $A_{pc} B_d=A_p B_s$ \citep[e.g.][]{cz99}. Thus, the surface area of the actual
polar cap $A_p=b^{-1} A_{pc}$, regardless of the actual shape of the polar cap. Consequently, one can write the polar
cap radius in the form $r_p=b^{-0.5}r_{pc}$. However, one should realize that this is only a characteristic dimension
of the actual polar cap. Indeed, the presence of strong surface magnetic field anomalies should make the shape of the
polar cap quite irregular.

For isolated neutron stars one might expect the surface to consist mainly of iron formed at the neutron star's birth
\citep{l01, um96}. Therefore, the charge depletion above the polar cap can result from bounding of the positive
$^{56}_{26}$Fe ions (at least partially) in the neutron star surface. If this is possible, then positive charges cannot
be supplied at the rate that would compensate the inertial outflow through the light cylinder. As a result, a
significant part of the unipolar potential drop (GJ69, RS75) develops above the polar cap, which can accelerate charged
particles to relativistic energies and power the pulsar radiation mechanism.

The cascading production of electron-positron plasma is crucial for limitation of growing gap potential drop
\citep[e.g. RS75,][ CR80 henceforth]{cr80}. The accelerated positrons will leave the acceleration region, while the
electrons will bombard the polar cap surface, causing a thermal ejection of ions. The ions will cause partial screening
of the potential drop, which can be described as $\Delta V=\eta({2\pi}/{cP})B_s h^2$, where $h$ is the height of the
acceleration region, $\eta=1-\rho_{i}/\rho_{\rm GJ}$ (CR80) is a shielding factor and $\rho_{i}$ is charge density of
ejected ions. The gap potential drop is completely screened when the total charge density $\rho=\rho_i+\rho_+$ reaches
the co-rotational value $\rho_{GJ}=e n_{GJ}$, where $n_{GJ}=1.4\times 10^{11}b\dot{P}_{-15}^{0.5}P^{-0.5}{\rm cm}^{-3}$
is the co-rotational charge number density .

GMG03 argued that the actual potential drop $\Delta V$ should be thermostatically regulated and the quasi-equilibrium
state should be established, in which heating due to electron bombardment is balanced by cooling due to thermal
radiation. The quasi-equilibrium condition is $Q_{cool}=Q_{heat}$, where $Q_{cool}=\sigma T_s^4$ is a cooling power
surface density by thermal radiation from the polar cap surface and $Q_{heat}=\gamma m_ec^3n$ is heating power surface
density due to back-flow bombardment, $\gamma=e\Delta V/m_ec^2$ is the Lorentz factor, $n=n_{GJ}-n_{i}=\eta n_{GJ}$ is
the number density of back-flowing plasma particles depositing their kinetic energy at the polar cap surface, $n_{i}$
is the charge number density of thermionic ions. It is straightforward to obtain an expression for the
quasi-equilibrium surface temperature in the form
\be
T_s=(6.2\times 10^4{\rm K})(\dot{P}_{-15}/{P})^{1/4}\eta^{1/2}b^{1/2}h^{1/2}.
\label{Ts}
\ee
Let us now interrelate the accelerating potential drop $\Delta V$ and the perpendicular (with respect of the magnetic
field lines) electric field $\Delta E$ which causes \EB\ drift. Following of RS75 we can argue that the tangent
electric field is strong only at the polar cap boundary where $\Delta E=0.5{\Delta V}/{h}=\eta({\pi}/{cP})B_sh$ (see
Appendix~A in GMG03 for details). Due to the \EB\ drift the discharge plasma performs a slow circumferential motion
with velocity $v_d=c\Delta E/B_s=\eta\pi h/P$. The time interval to make one full revolution around the polar cap
boundary is $\hat{P}_3\approx 2\pi r_p/v_d$. One then has
\be
\frac{\hat{P}_3}{P}=\frac{r_p}{2\eta h}.
\label{P3P}
\ee
If the plasma above the polar cap is fragmented into filaments (sparks) which determine the intensity structure of the
instantaneous pulsar radio beam, then in principle, the tertiary periodicity $\hat{P}_3$ can be measured/estimated from
the pattern of the observed drifting subpulses \citep[e.g.][]{dr99, gs03}. According to RS75, $\hat{P}_3=NP_3$, where
$N$ is the number of sparks contributing to the drifting subpulse phenomenon observed in a given pulsar and $P_3$ is
the primary drift periodicity (distance between the observed subpulse drift bands). On the other hand $N\approx 2\pi
r_p/2h$ (GS00). Thus, one can write the shielding factor in the form $\eta\approx (1/2\pi)(P/P_3)$ , which depends only
on an easy-to-measure primary drift periodicity. Apparently, the shielding parameter $\eta$ should be much smaller than
unity.

The X-ray thermal luminosity is $L_x=\sigma T_s^4\pi r_p^2=1.2\times 10^{32}(\dot{P}_{-15}/P^3)(\eta h/r_p)^2$~erg/s,
which can be compared with the spin-down power $\dot{E}=I\Omega\dot{\Omega}=3.95 I_{45}\times
10^{31}\dot{P}_{-15}/P^3$~erg/s, where $I=I_{45}10^{45}$g\ cm$^2$ is the neutron star moment of inertia (bellow we
assume that $I_{45}=1$). Using equation~(\ref{P3P}) we can derive the thermal X-ray luminosity and its efficiency as
$L_x=2.5\times 10^{31}(\dot{P}_{-15}/P^3)(P/\hat{P}_3)^{2}$, or in the simpler form representing the efficiency with
respect to the spin-down power
\be \frac{L_x}{\dot{E}}=0.63 \left(\frac{P}{\hat{P}_3}\right)^{2}
\label{Lx},
\ee
which is useful for comparison with observations. One should realize that this equation holds only for thermal X-rays
from hot spot and cannot be applied neither to cooler radiation from the entire stellar surface nor to the
magnetospheric component.

We can see that $L_x$ in these equations depends only on radio observables. It is particularly interesting and
important that this equations does not depend explicitly on $\eta, b$ and $h$, but only on their combination expressed
by Eq.(\ref{P3P}). Although one has to be careful whether all our assumptions are satisfied in real pulsars, it seems
that we have found useful, relatively easily testable relationship between the properties of drifting subpulses
observed in radio band and the characteristics of thermal X-ray emission from the polar cap heated by sparks associated
with these subpulses.

Using Eq.~(\ref{P3P}) we can write the polar cap temperature in the form
\be
T_s=(5.1\times 10^6 {\rm
K})b^{1/4}\dot{P}^{1/4}_{-15}P^{-1/2}\left(\frac{\hat{P}_3}{P}\right)^{-1/2}
\label{Ts4},
\ee
where the enhancement coefficient $b=B_s/B_d\approx A_{pc}/A_{bol}$, $A_{pc}=\pi r^2_{pc}$ and $A_{bol}=A_p$ is the
actual emitting surface area (bolometric). Since $A_{bol}$ can be determined from the black-body fit to the spectrum of
the observed hot-spot thermal X-ray emission, the above equation, similarly to Eq.(\ref{Lx}), depends only on combined
radio and X-ray data.

 Although Eq.(\ref{Lx}) seems generic in any IAR model in which drift rate and the heating rate are determined by the
same electric field (including pure vacuum gap of RS75), Eqs. (\ref{Lx}) and (\ref{Ts4}) considered together describe
exclusively PSG model. The observational tests of this model are discussed in section 3 and summarized in Table 1. For
PSRs B0943$+$10 and B1133+16, which are the only two pulsars in which both $\hat{P}_3$ as well as $L_x$ and $A_{bol}$
are measured, the above equation seems to hold quite well. Interestingly, in both cases $L_x/\dot{E} \sim 10^{-3}$ and
$T_s \sim 3$ MK.

\section{Comparison with observational data}

Table 1 presents the data for two pulsars, which we believe show clear evidence of thermal X-ray emission from the
polar caps as well as they have known values of tertiary subpulse drift periodicity. The predicted value of $\hat{P}_3$
and/or $L_x$ were computed from Eq.(\ref{Lx}), while the predicted values of $T_s$ were computed from Eq.(\ref{Ts4}),
with $b=A_{pc}/A_{bol}$ determined observationally.

{\it\bf PSR B0943$+$10.} This is the best studied drifting subpulse radio pulsars with $P=1.09$~s,
$\dot{P}_{-15}=3.52$, $\dot{E}=10^{32}\ {\rm erg\ s}^{-1}$, $P_3=1.86P$, $\hat{P}_3=37.4P$ and $N=\hat{P_3}/P_3=20$
\citep{dr99}. It was observed by ZSP05, who obtained an acceptable thermal BB fit with bolometric luminosity
$L_x=(5^{+0.6}_{-1.6})\times 10^{28}\ {\rm erg\ s}^{-1}$ and thus $L_x/\dot{E}=(0.49^{+0.06}_{-0.16})\times 10^{-3}$.
The bolometric polar cap surface area $A_{bol}=10^7[T_s/(3\times 10^6 {\rm K})]^{-4} {\rm
cm}^2\sim(1^{+4.0}_{-0.4})\times 10^7\ {\rm cm}^2$ is much smaller than the conventional polar cap area $A_{pc}=6\times
10^8\ {\rm cm}^2$. This all correspond to the best fit temperature $T_s \sim 3.1\times 10^6$~K (see Fig.~1 in ZSP05).
The predicted value of $L_x/\dot{E}$ calculated from Eq.~(\ref{Lx}) agrees very well with the observational data. The
surface temperature $T_s$ calculated from Eq.~(\ref{Ts4}) with $b=A_{pc}/A_{bol}$ is also in good agreement with the
best fit. The shielding factor $\eta=0.09$.

{\it\bf\ PSR B1133$+$16.} This pulsar with $P=1.19$~s, $\dot{P}_{-15}=3.7$, and $\dot{E}=9\times 10^{31}\ {\rm erg\
s}^{-1}$ is almost a twin of PSR B0943$+$10. KPG06 observed this pulsar with Chandra and found an acceptable BB fit
$L_x/\dot{E}=(0.77^{+0.13}_{-0.15})\times 10^{-3}$, $A_{bol}=(0.5^{+0.5}_{-0.3})\times 10^7~{\rm cm}^2$ and $T_s\approx
2.8\times 10^6$~K. These values are also very close to those of PSR B0943$+$10, as should be expected for twins. Using
equation (\ref{Lx}) we can predict $\hat{P}_3/P=27^{+5}_{-2}$ for B1133+16. Interestingly, \citet{n96} obtained
fluctuation spectrum for this pulsar with clearly detected long period feature corresponding to about $32P$. Most
recently, WES06 found $P_3/P=3\pm 2$ and long period feature corresponding to $(33\pm 3) P $ in the fluctuation
spectrum of PSR B1133+16. The latter value seem to coincide with that of \citet{n96}, as well as with our predicted
range of $\hat{P}_3$. We therefore claim that this is the actual tertiary periodicity in PSR B1133+16 and show it in
parenthesis in Table 1.

It is worth noting that $\hat{P}_3/P_3=33\pm 3$ is quite close to 37.4 measured in the radio twin PSR B0943+10. Note
also that the number of sparks predicted from our hypothesis is $N=\hat{P}_3/P_3=(33\pm 3)/(3\pm 2)=11^{+25}_{-6}$, so
it can also be close to 20, as in the case of twin PSR B0943$+$10. The shielding factor $\eta=0.05^{+0.11}_{-0.02}$.

\section{Conclusions and Discussion}

Within the PSG model of IAR developed by GMG03 we derived a simple relationship between the X-ray luminosity $L_x$ from
the polar cap heated by sparks and the tertiary periodicity $\hat{P}_3$  of the subpulse drift observed in radio band.
In PSRs B0943$+$10 and B1133+16 for which both $L_x$ and $\hat{P}_3$ are known, the predicted relationship between
observational quantities holds well enough.

Both the heating and the drifting rate depends on the gap potential drop. In this paper we made the point that there is
continuum of cases between pure vacuum gap and the space charge limited flow. The original RS75 model predicts much too
high a subpulse drift rate and an X-ray luminosity. Other available acceleration models predict too low a luminosity
and the explanation of drifting subpulse phenomenon is generally not clear (see ZSP05 for more detailed discussion).
Approximately, the bolometric X-ray luminosity for the space charge limited flow is about $(10^{-4} \div
10^{-5})\dot{E}$ \citep{hm02}, and for the pure vacuum gap (RS75) is about $(10^{-1} \div 10^{-2})\dot{E}$ (ZSP05),
while for the PSG (GMG03) is $\sim 10^{-3}\dot{E}$ (this paper). The latter model also predicts right \EB\ plasma drift
rate. Thus, combined radio and X-ray data are consistent only with the PSG model, which requires very strong (generally
non-dipolar) surface magnetic fields. Observations of the hot-spot thermal radiation almost always indicate bolometric
polar cap radius much smaller than the canonical value (at least by an order of magnitude). Most probably such a
significant reduction of the polar cap size is caused by the flux conservation of the non-dipolar surface magnetic
fields connecting with the open dipolar magnetic field lines at distances much larger than the neutron star radius.
Part of this reduction can follow from the fact that discrete sparks do not heat an entire polar cap area (but
presumably more than a half), but uncertainties in determination the size of the polar cap are likely to account for
this effect.

\section*{Acknowledgments}
We acknowledge support from the Polish State Committee for scientific research under Grant 2 P03D 029 26.

{}

\end{document}